\documentstyle[twocolumn,seceq,epsf]{jpsj}

\def\runtitle{ Neutron Diffraction Study of the Irreversible R-M$_{\rm
    A}$-M$_{\rm C}$ }

\def\runauthor{ Kenji {\sc Ohwada} {\it et al.} }

\title{ Neutron Diffraction Study of the Irreversible R-M$_{\rm
    A}$-M$_{\rm C}$ Phase Transition in Single Crystal
  Pb[(Zn$_{1/3}$Nb$_{2/3}$)$_{1-x}$Ti$_{x}$]O$_{3}$}

\author{ Kenji {\sc Ohwada}\footnote{Present address: SPring-8, JAERI, 
    1-1-1 Kouto Mikazuki-cho Sayo-gun Hyogo
    679-5148, Japan }, Kazuma {\sc Hirota}$^{1}$, Paul W. {\sc
    Rehrig}$^{2}$, Peter M. {\sc Gehring}$^{3}$, Beatriz {\sc
    Noheda}$^{4}$, Yasuhiko {\sc Fujii}, Seung-Eek Eagle{\sc Park}$^{5}$,
    and Gen {\sc Shirane}$^{4}$ }

\inst{ 
  Neutron Scattering Laboratory, ISSP, University of Tokyo, 
  106-1 Shirakata, Tokai, Ibaraki 319-1106, Japan \\ 
  $^1$Department of Physics, Tohoku University, Sendai 980-8578, Japan \\ 
  $^2$Materials Research Laboratory, The Pennsylvania State
  University, PA 16802, U.S.A. \\ 
  $^3$NIST Center for Neutron Research, NIST, Gaithersburg, 
  MD 20899-8562, U.S.A. \\ 
  $^4$Department of Physics, Brookhaven National Laboratory, 
  Upton, NY 11973-5000, U.S.A. \\   
  $^5$Fraunhofer-IBMT Technology Center Hialeah, Hialeah, FL 33010, U.S.A.\\
}

\recdate{ \today }

\abst{ Single crystals of the relaxor PZN-$x$PT display an enormously
  strong piezoelectric character.  Recent x-ray scattering studies
  have revealed novel electric-field induced phase transitions in
  PZN-8\%PT.  As-grown crystals exhibit a rhombohedral structure that,
  under application of an electric field oriented along [001],
  transforms into a monoclinic (M$_{\rm A}$) phase, and then
  irreversibly to another monoclinic (M$_{\rm C}$) phase with
  increasing field strength. Since the latter phase change is very
  unusual, its transition sequence has been investigated by using
  triple-axis neutron scattering techniques so that the ``skin
  effect'' observed by x-ray scattering can be avoided, and the entire
  crystal bulk is probed.  Contour maps of the elastic scattering have
  been mapped out in each phase in the (HOL) zone with high
  $q$-resolution.  Increasing the field strength within the M$_{\rm
    C}$ phase induces a sharp $c$-axis jump around 15 kV/cm.  This
  jump was observed easily with x-rays in previous studies, but it was
  not observed in 5 different crystals examined with neutrons.  A
  subsequent high-energy x-ray study of the same crystals showed that
  the $c$-axis jump is distributed within the crystal volume, thereby
  washing out the jump.  The observed R-M$_{\rm A}$-M$_{\rm C}$ 
  transformational path is in perfect accord with very recent first 
  principles calculations by Bellaiche, Garcia, and Vanderbilt 
  in the PZT system.}

\kword{relaxor, ferroelectric, PZN-$x$PT, phase transition, electric
  field, neutron scattering}

\begin{document}
\sloppy
\maketitle

\section{Introduction}

Several recent experimental and theoretical studies have significantly
advanced our understanding of the origin of the exceptionally high
piezoelectric and dielectric responses that have been reported in the
lead-oxide class of relaxor ferroelectrics.  The x-ray work by Noheda
{\it et al.}~\cite{Noheda1,Noheda2} uncovered a previously unknown
sliver of monoclinic (M) phase nestled against the morphotropic phase
boundary (MPB) separating the rhombohedral (R) and tetragonal (T)
regions of the phase diagram of Pb(Zr$_x$Ti$_{1-x}$)O$_3$
(PZT)~\cite{Jaffe}, a perovskite system that is the material of choice
in the fabrication of high performance actuators and transducers for
industrial applications.  Subsequent x-ray measurements have revealed
the presence of a similar narrow region of monoclinic phase in the
relaxor Pb[(Zn$_{1/3}$Nb$_{2/3}$)$_{1-x}$Ti$_x$]O$_3$
(PZN-$x$PT)~\cite{Noheda3,Cox}, a ferroelectric material that exhibits
an ultra-high strain in high field fully 10 times that of
PZT~\cite{Kuwata1,Kuwata2,Park1}.  In both PZN-$x$PT and PZT the
maximum piezoelectric activity is located on the R side very near the
MPB, thereby underscoring the important role played by the monoclinic
phase.

These observations triggered the theoretical work of Fu and Cohen who
introduced the concept of the polarization rotation mechanism to
explain the ultra-high electromechanical response \cite{Fu}.  Whereas
in the conventional ferroelectric phases of tetragonal PbTiO$_3$ and
BaTiO$_3$ the polarization vector points along the [001] and [111]
directions, respectively, the monoclinic symmetry afforded the PZT and
PZN-$x$PT compounds near the MPB allows the polarization vector a much
greater degree of freedom as it is only constrained to lie within a
monoclinic plane defined by the pseudo-cubic [001] and [111] directions
in the case of PZT, i.e.,  the (1$\bar{1}$0) plane, and the [001]
and [101] direction in the case of PZN-$x$PT, i.e., the (001) plane.
In the monoclinic phase, the polarization 
direction can easily adjust to the electric field, which naturally
results in a large piezoelectric response.  A polarization rotation
path within the (1$\bar{1}$0) plane can be 
considered phenomenologically by expanding the Devonshire theory of
ferroelectrics to eighth or higher order as done by Vanderbilt and
Cohen~\cite{Cohen}.  Indeed, such an expansion allows for monoclinic
phases in addition to R, T and orthorhombic (O) phases, 
and is able to generate a global phase diagram within eighth order.

PZN-$x$PT is a perovskite ($ABO_3$) solid solution of the relaxor
Pb(Zn$_{1/3}$Nb$_{2/3}$)O$_3$ ($T_c$ = 410~K) and the ferroelectric
PbTiO$_3$ ($T_c$ = 673~K).  The system has cubic symmetry at high
temperatures and undergoes a diffuse ferroelectric phase transition
above room temperature for low values of $x$~\cite{Kuwata1}, which is
strongly frequency-dependent (hence the term ``relaxor'').  This
unusual diffuse phase transition is believed to be due to the complex
mixture of Zn$^{2+}$, Nb$^{5+}$, and Ti$^{4+}$ cations which give the
perovskite $B$-site a marked mixed valence character.  Unlike PZT,
PZN-$x$PT can be grown in single crystal form which makes it
possible to study their structural properties in detail.

%
%
\begin{figure}[t]
 \begin{center}
      \epsfxsize=8.5cm
      \epsfbox{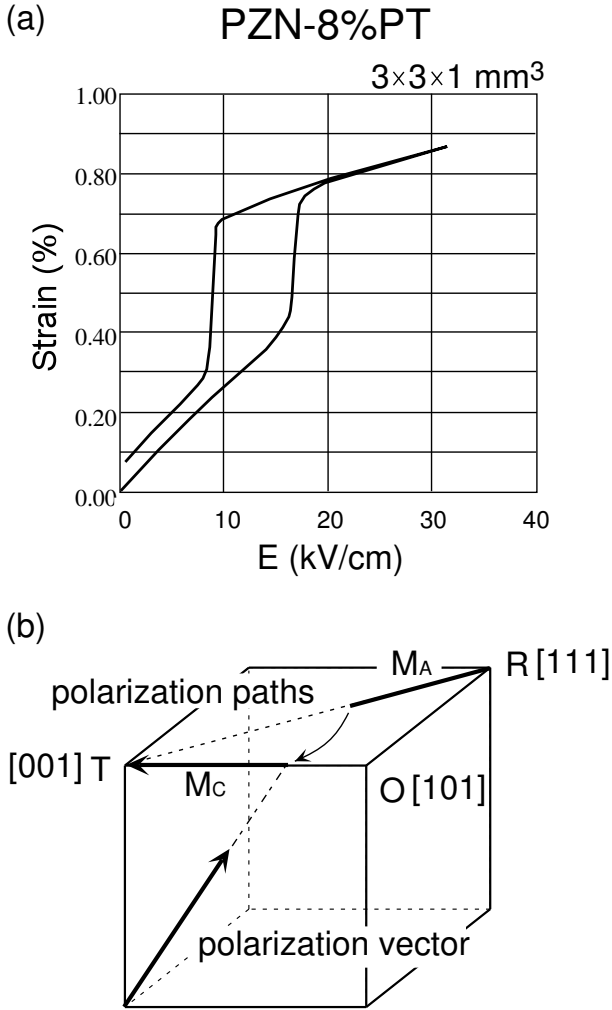}
 \end{center}
\caption{ (a) Electric field dependence of the strain measured on the 
  3$\times$3$\times$1 mm$^3$ PZN-8\%PT crystal (Ref.~\cite{Park1}).
  (b) Polarization path for the PZN-8\%PT crystal.}
\label{fig:1}
\end{figure}  

Single crystals of PZN-$x$PT exhibit the greatest piezoelectric
response for $x$=8\%, which is located near the MPB on the R
side~\cite{Park1}.  Strain field loops for PZN-8\%PT (see
Fig.~\ref{fig:1}(a)) and PZN-4.5\%PT, where the electric field was
oriented along the pseudo-cubic [001] direction, show that the strain
increases linearly below a threshold field, and then suddenly
jumps~\cite{Park1}, an effect which is known as a $c$-axis jump.  This
threshold field, as well as the piezoelectric response, decreases as
$x$ increases beyond the MPB and into the tetragonal phase.  X-ray
scattering experiments on PZN-$8$\%PT by Durbin {\it et al.} were able
to reproduce this $c$-axis jump behavior~\cite{Durbin1}.  In addition,
Durbin {\it et al.} discovered that an irreversible change in phase
takes place from the ``as-grown'' state to a different phase with the
application of an electric field~\cite{Durbin2}.  Subsequent x-ray
measurements by Noheda {\it et al.} identified the proper symmetries
of the various phases~\cite{Noheda3}.  In particular, Noheda {\it et
  al.} showed that poled crystals have a monoclinic (M$_{\rm{C}}$)
symmetry that is different from that found in the PZT system
(M$_{\rm{A}}$)~\cite{Noheda3}.  Furthermore, it was shown that
as-grown PZN-$x$PT crystals have a rhombohedral symmetry that
transforms irreversibly upon application of an electric field along
[001] to a monoclinic (M$_{\rm{C}}$) phase via an intermediate
monoclinic (M$_{\rm{A}}$) phase~\cite{Noheda3}.  This sequence is
surprising as it runs counter to the theoretical sequence proposed by
Fu and Cohen of R-M$_{A}$-T based on calculations performed for
BaTiO$_3$~\cite{Fu}.  Instead, it appears that the M$_{\rm{A}}$ phase
is needed for the R-M$_{\rm{C}}$ phase transition~\cite{Noheda3}.
Figure~\ref{fig:1}(b) depicts the path followed by the polarization
rotation for the R-M$_{\rm{A}}$-M$_{\rm{C}}$ phase sequence 
proposed~\cite{Noheda3}. These results are in excellent agreement 
with the first principles calculations done for the PZT system 
by Bellaiche et al.~\cite{Bellaiche}, which were made public during 
the preparation of this manuscript.

It is now clear that PZN-8\%PT shows two novel field-induced
phase transitions.  To clarify the origin of the exceptional 
piezoelectric character of this system, 
we have studied this complicated transformation sequence in detail.
Discrepancies between
recent x-ray and neutron scattering studies suggest that the
near-surface region (or $skin$) of a single crystal of PZN-$x$PT may
behave differently from the crystal bulk.  We have thus carried out
neutron scattering experiments under an applied electric field where
we have made significant efforts to improve the instrumental $q$
resolution.

%
%

\begin{fullfigure}[t]
 \begin{center}
     \epsfxsize=15cm
     \epsfbox{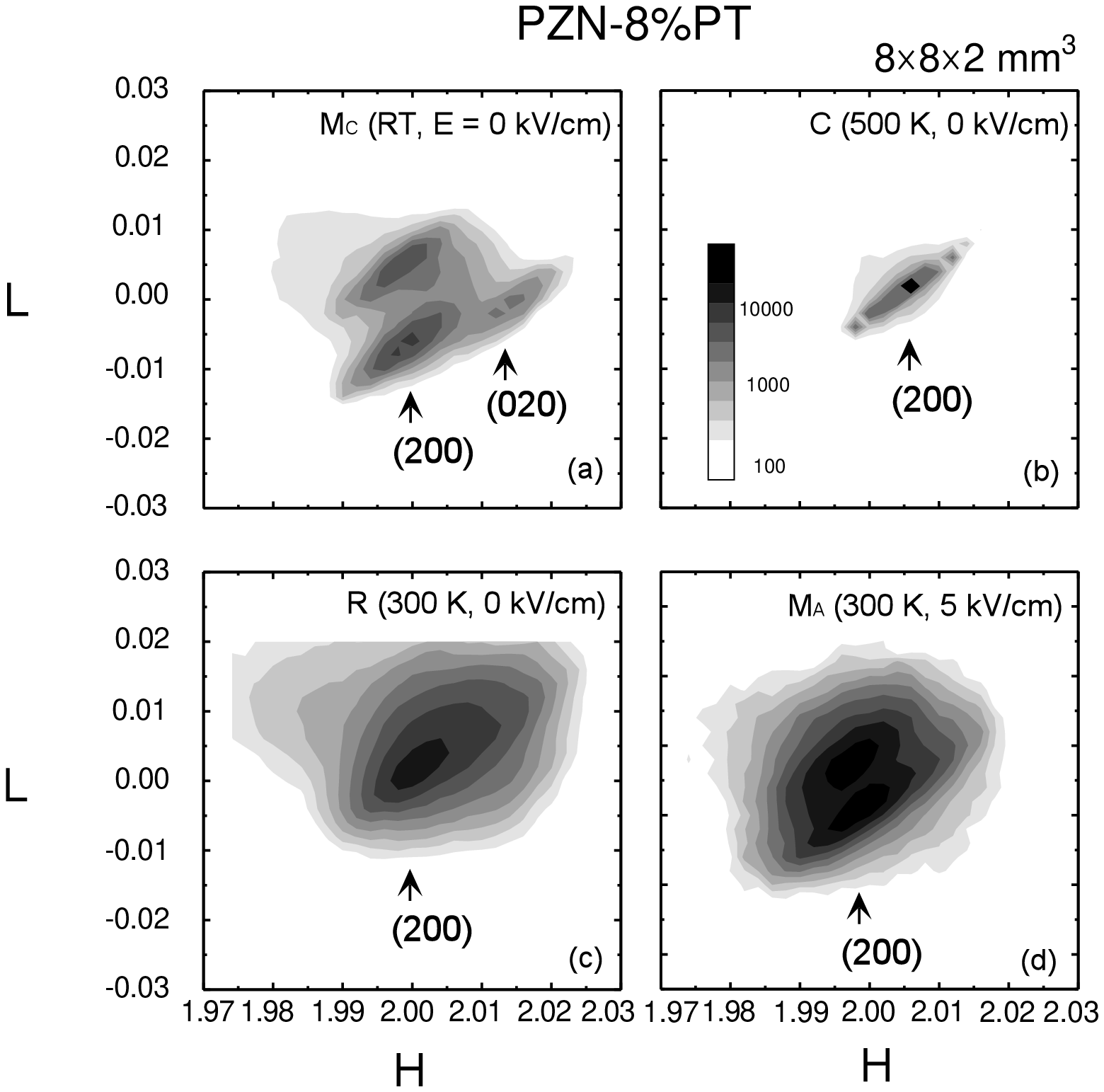}
 \end{center}
\caption{Contour maps of the (HOL) zone around (200) showing
  the sequence of phase transitions observed in the PZN-8\%PT
  (8$\times$8$\times$2 mm$^{3}$ crystal as a function of temperature
  and electric field.  (a) M$_{\rm{C}}$-phase at RT and E = 0 kV/cm,
  (b) C-phase at T = 500 K and E = 0 kV/cm, (c) R-phase at T = 300 K
  and E = 0 kV/cm, (d) M$_{\rm{A}}$-phase at T = 300 K and E = 5
  kV/cm.}
\label{fig:2}
\end{fullfigure}

\section{Experimental Details }

Neutron scattering measurements have been performed on 5 different
single crystals of 8.0\%PT having dimensions
2$\times$2$\times$2~mm$^3$ (two of them), 3$\times$3$\times$1~mm$^3$,
6$\times$6$\times$3~mm$^3$, and 8$\times$8$\times$2~mm$^3$.  Similar
measurements were also performed on 2 single crystals of 4.5\%PT with
dimensions 2$\times$2$\times$0.5~mm$^3$, and
3$\times$3$\times$1~mm$^3$.  The two 2$\times$2$\times$2~mm$^3$
crystals were also used in the first x-ray diffraction study of
8.0\%PT by Durbin {\it et al}~\cite{Durbin1}.  These crystals were
grown at Penn State University, and the strain curves measured along
[001] as a function of electric field strength were always used as a
gauge of crystal quality.  As shown in Fig.~1, both crystals exhibit
sharp jumps in the $c$-axis lattice spacing around 15~kV/cm.

The neutron diffraction experiments were carried out at the JRR-3M
reactor located at the Japan Atomic Energy Research Institute (JAERI)
in Tokai.  Additional measurements were performed at the NBSR reactor
located at the NIST Center for Neutron Research in Gaithersburg,
Maryland.  A very high instrumental $q$-resolution of 0.003 \AA$^{-1}$
full-width half maximum (FWHM) was achieved by using a perfect Ge
(220) crystal as analyzer.  The (220) Bragg planes of Ge have a
$d$-spacing of 2.0~\AA, a value that closely matches that of the (200)
Bragg planes of the PZN-$x$PT perovskite oxides.  This technique of
lattice matching results in a high $q$-resolution.  The same method
was used in prior high resolution studies of CuGeO$_3$~\cite{Hirota}.

A supplemental high-energy x-ray study was performed at the
superconducting-wiggler beam line X17B located at the National
Synchrotron Light Source (NSLS) at Brookhaven National Laboratory in
Upton, New York.  X-rays of this energy can pass through a 3 mm thickness
of PZN and permits the study of cross sections as small as 50 microns .

\section{The Irreversible Structural Phase Transformation Sequence 
  R-M$_{\rm{A}}$-M$_{\rm{C}}$ }
%
%
\begin{figure}[t]
 \begin{center}
     \epsfxsize=8cm
     \epsfbox{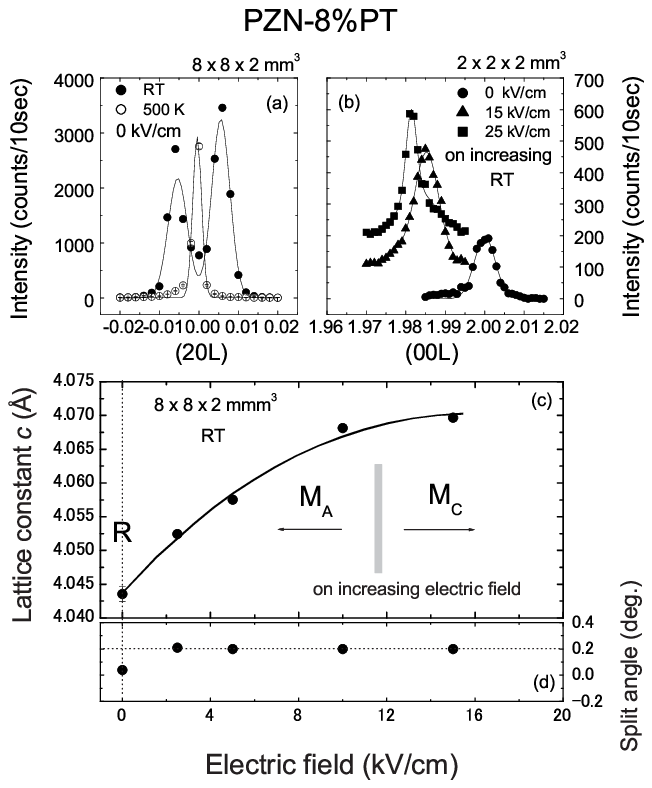}
 \end{center}
\caption{(a) (20L) peak profiles in the M$_{\rm{C}}$ (solid circles) 
  and C (open circles) phases of the PZN-8\%PT
  8$\times$8$\times$2 mm$^3$ crystal.  (b) Electric field dependence
  of the (00L) peak profiles taken at several fields 0, 15, 25 kV/cm.
  (c) Lattice constant $c$ and (d) peak split angle as a function of
  electric field taken at the R-M$_{\rm{A}}$-M$_{\rm{C}}$ transition
  sequence.  Solid lines drawn through the data points are guides to
  the eyes.}
\label{fig:3}
\end{figure}

To study the R-M$_{\rm{A}}$-M$_{\rm{C}}$ phase transition sequence,
contour maps of the elastic scattering have been mapped out in each
phase in the (HOL) zone around the (200) reciprocal lattice position
with high $q$-resolution as a function of electric field applied along
the [001] direction.  These data are shown in Fig.~\ref{fig:2}, and
were taken on an 8$\times$8$\times$2 mm$^{3}$ PZN-8\%PT crystal that
had been poled in a field of 15 kV/cm along [001] (i.e., normal to
the 8$\times$8 mm$^{2}$ crystal surface).  The contour maps provide an
extremely useful two-dimensional view of the scattering associated
with each phase.  The scattering intensity has been represented on a
logarithmic gray scale as shown in the inset of Fig.~\ref{fig:2}(b).
Our measurements began with the poled crystal at room temperature (RT)
and in zero applied field.  In this state, a domain structure is found
in which the (200) peak is split and the (020) peak is clearly
observed (see Fig.~\ref{fig:2}(a)).  The $a$ and $c$ lattice constants
are nearly equal, but the $b$-axis spacing is noticeably smaller.
This suggests that the polarization direction lies in the $ac$-plane,
and that the crystal is indeed in the M$_{\rm{C}}$-phase as a result
of the poling.  The data shown in Fig.~\ref{fig:2}(a) are consistent
with the x-ray results on PZN-8\%PT obtained by Noheda {\it et
  al}~\cite{Noheda3}.  Next, the crystal was warmed above 500 K to
restore it to the cubic (C) phase.  The result is a single sharp peak
near (200) (slightly shifted due to the thermal expansion) as shown in
Fig.~\ref{fig:2}(b).  The (20L) scattering profile for each of these
two phases is shown in Fig.~\ref{fig:3}(a) for comparison.  The cubic
phase profile is exceedingly sharp, reflecting the relaxation of the
local strains by thermal fluctuations.  The crystal was then cooled to
RT, whereupon the phase becomes R, as is observed in other crystals in
their as-grown state at RT (Fig.~\ref{fig:2}(c)).  The substantially
broadened peak width suggests the generation of local distortions
caused by the randomly-oriented $\langle$111$\rangle$ polar nanoregions (PNR).

An electric field was first applied to the crystal in the R phase,
after cooling from 500~K to RT.  
From this point on, all data were
taken at room temperature.  Immediately after E = 0 kV/cm, 
the rhombohedral phase transformed into the M$_{\rm{A}}$ phase.  
This is shown in Fig.~\ref{fig:2}(d) for E =
5~kV/cm.  We also display the electric field dependence of the
$c$-axis lattice constant and the observed monoclinic angle in
Fig.~\ref{fig:3}(c) and (d).  From Fig.~\ref{fig:2}(c) we see that the
(200) peak splits into two peaks at 5~kV/cm, however no other peaks
appear.  This suggests that the polarization direction is oriented
between $\langle$111$\rangle$ and $\langle$001$\rangle$, 
and thus corresponds to the M$_{\rm{A}}$-phase.
The angle formed by the two split peaks is about 0.2$^{\rm o}$ (see
Fig.~\ref{fig:3}(d)) and is independent of electric field in the
M$_{\rm{A}}$-phase.  The appearance of the (020) peak signals the
M$_{\rm{A}}$-M$_{\rm{C}}$ phase transition, and is attained between E =
10 and 15~kV/cm.  The M$_{\rm{C}}$ phase is retained as the
ground state structure of the PZN-8\%PT crystal, even after removal of
the applied field.  This is demonstrated by Fig.~\ref{fig:2}(a) which
depicts the state of the crystal after poling along [001] in a field
of 15~kV/cm.  The R phase, by contrast, is not recovered at E =
0~kV/cm, hence the transition to the M$_{\rm{C}}$ phase is irreversible.

We examine the electric field dependence of the $c$-axis lattice
constant more closely in a second crystal (2$\times$2$\times$2 mm$^3$),
poled such that it is in the M$_{\rm{C}}$ phase, in the next section.

%
%

\begin{figure}[t]
 \begin{center}
     \epsfxsize=8.5cm
     \epsfbox{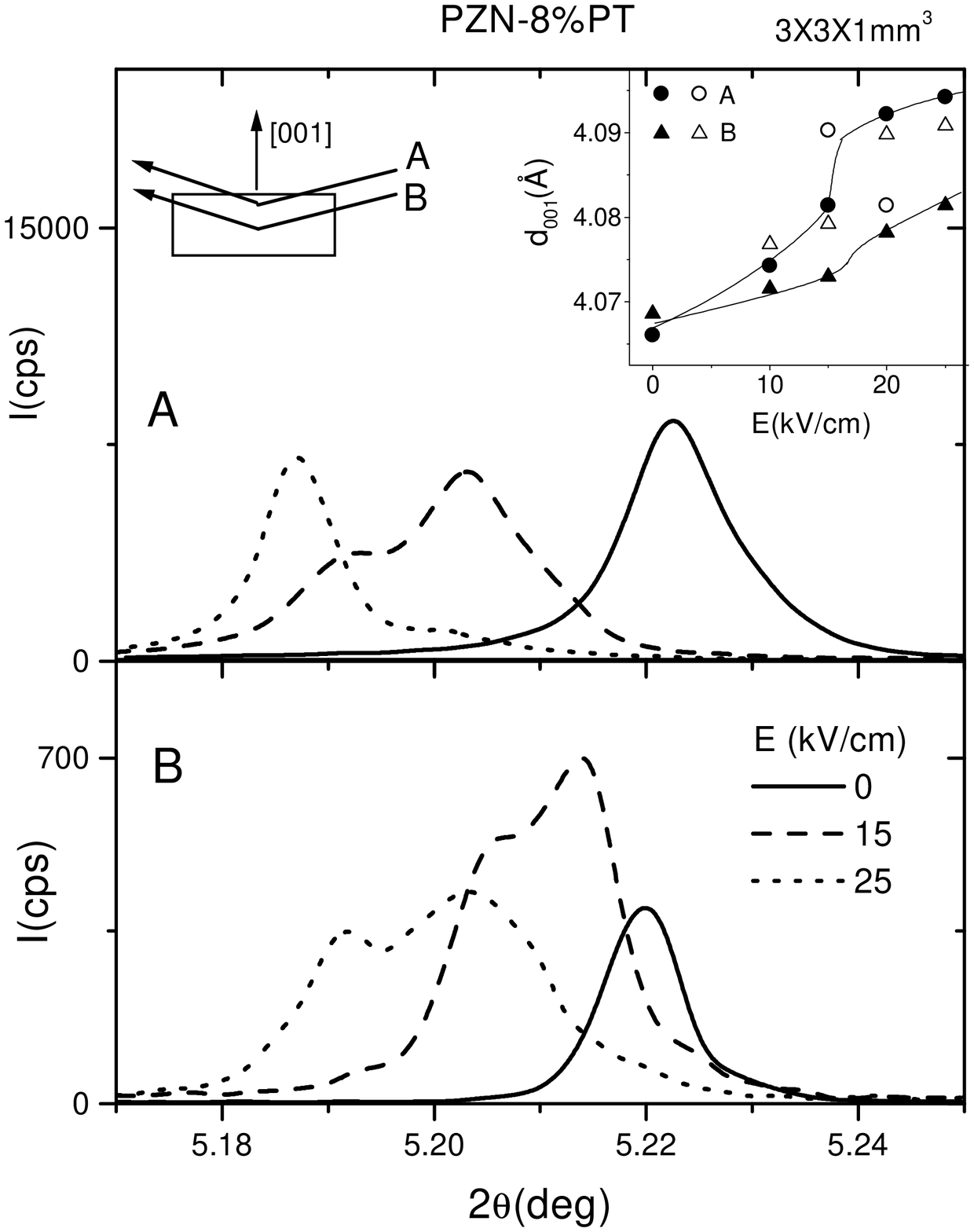}
 \end{center}
\caption{High-energy x-ray results on PZN-8\%PT crystals showing 
  the $c$-axis jump is distributed within the crystal volume.
  Diffraction patterns resulting mainly from the surface (bulk) are
  shown in the upper (lower) panel. 
  The evolution of the spacing of the 001 plane d$_{001}$
  with the electric field is shown 
  in the inset for both surface (A) and the bulk (B). 
  The major (minor) component in the diffraction peaks is represented
  as solid (open) symbols.}
\label{fig:4}
\end{figure}  

\section{The $c$-axis Jump}

The search for the $c$-axis jump using neutron scattering techniques
grew out of our initial attempts to study the effects of an applied
electric field on the soft phonon anomalies reported in PZN-8\%PT by
Gehring {\it et al.} \cite{Gehring}.  During these measurements it was
noted that the $c$-axis jump observed with x-rays by Durbin {\it et
  al.} \cite{Durbin1} was missing.  The magnitude of the jump in the
$c$-axis spacing is of order 0.5\%, and so should have been clearly
visible with neutron scattering techniques.  The discrepancy between
neutron and x-ray scattering results stimulated a much more intensive
search for the $c$-axis jump, and involved the study of a total of 5
different single crystals, including two 2$\times$2$\times$2~mm$^3$
crystals that were used in the x-ray study.  Nevertheless the $c$-axis
jump was not reproduced in the neutron scattering data.  The absence
of the $c$-axis jump in the neutron experiments suggested that the
outer volume, or ``skin,'' of the crystals could be behaving
differently from the crystal interior given that the x-rays and
neutrons probe different volumes of the crystal.  The x-rays used by
Durbin {\it et al.} were obtained from a Cu $K_{\alpha}$ rotating
anode source with a wavelength $\lambda$ = 1.541~\AA (8 keV), and
penetrate only of order 1~$\mu$m (1~$\mu$m = 10,000~\AA) into the
crystal due to the high lead content of these samples.  Neutrons, by
contrast, probe the entire crystal volume.  This idea prompted a
subsequent x-ray study by Noheda {\it et al.} using much higher energy
x-rays (67 keV) in which it was shown that so-called ``skin effects''
are in fact important in these lead-oxide relaxor
systems~\cite{Noheda3}.  Figure~\ref{fig:4} shows the results obtained
using high-energy x-rays on crystals of PZN-8\%PT.  The scattering
that occurs from mainly the surface (upper panel) and the bulk (lower
panel) reveal that the sharp $c$-axis jump is in fact distributed over
the crystal volume, thereby effectively washing out the $c$-axis jump.
Motivated by this x-ray work, we carried out a new series of neutron
scattering measurements under an applied electric field with a
significantly higher $q$-resolution than before.

%
%
\begin{figure}[t]
 \begin{center}
     \epsfxsize=8.5cm
     \epsfbox{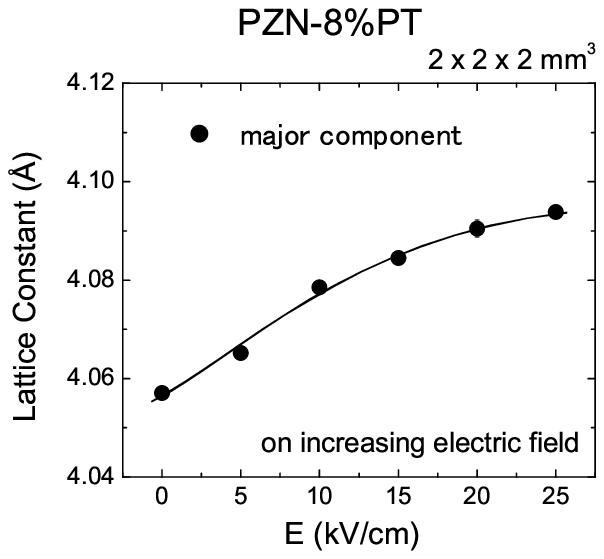}
 \end{center}
\caption{Electric field dependence of the $c$-axis lattice constant 
  of PZN-8\%PT measured with increasing field.  The solid circles
  represent the major component of the peak profiles. The solid line
  is a guide for the eyes. There is no sharp jump as was seen in
  Fig.~\ref{fig:1}(a).}
\label{fig:5}
\end{figure}  

\begin{figure}[t]
 \begin{center}
     \epsfxsize=8.5cm
     \epsfbox{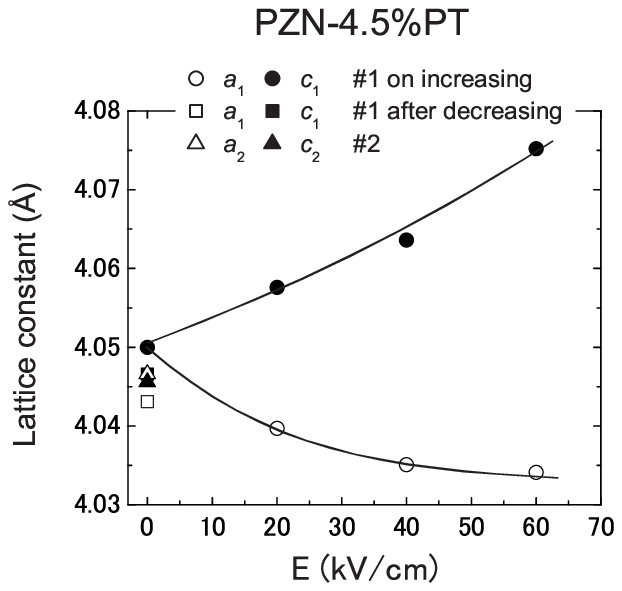}
 \end{center}
\caption{Electric field dependence of the lattice constant $a$ and $c$ 
  of PZN-4.5\%PT measured on increasing field. The solid line is
  guided for the eyes. There is also no jump as reported in
  Ref~\cite{Park2}}
\label{fig:6}
\end{figure}

Selected (00L) peak profiles at E = 0.0, 15.0 and 25.0~kV/cm are shown
in Fig.~\ref{fig:3}(b).  The important feature to note is the asymmetry
of the peak profiles due to the shoulder exhibited on the low-$q$ side
at E = 15~kV/cm, and then on the high-$q$ side at E = 25~kV/cm.  This
asymmetry is a measure of the strain distribution present throughout
the crystal because neutrons probe the entire crystal volume.  More
importantly, this asymmetry demonstrates that the strain is not
uniform within the crystal.  In addition, the peak profile sharpens
with increasing field.  These results are consistent with the previous
high-energy x-ray results of Noheda {\it et al.,}~\cite{Noheda3} and
confirm the presence of the distributed phase transition.  By
extracting the peak position of the major component of the measured
profiles, one obtains the field dependence of the $c$-axis lattice
constant shown in Fig.~\ref{fig:5}.  The $c$-axis gradually expands and
shows no jump around the threshold field of 16~kV/cm.  For comparison,
the experimental results obtained at the NIST Center for Neutron
Research on a single crystal of PZN-4.5\%PT are shown in
Fig.~\ref{fig:6}.  Again, there is no jump at its threshold field of
35~kV/cm~\cite{Park2}.  These results indicate that the strain is
distributed throughout the crystal volume, and thus provides a natural
explanation for why neutron scattering, which is a bulk probe, does
not observe a sharp $c$-axis jump.

%
%

\section{Conclusion}

Neutron scattering experiments have been carried out with very sharp
$q$-resolution on single crystals of the highly piezoelectric relaxor
material PZN-$x$PT as a function of applied electric field.  Contour
plots of the elastic scattering measured in the (HOL) zone near the
(200) Bragg peak in PZN-8\%PT confirm the irreversible
R-M$_{\rm{A}}$- M$_{\rm{C}}$ sequence of transformations first
reported in the x-ray studies of Durbin and
Noheda~\cite{Durbin1,Noheda3}.  While the presence of a monoclinic
phase in a cubic perovskite ferroelectric system is quite unusual, it
can now be understood within the framework of an extended Devonshire
theory for strongly anharmonic crystals for which higher order terms
become important.\cite{Cohen} However, the actual path taken by the
polarization in PZN-8\%PT under the influence of an applied electric
field was unexpected.  The first principles
calculations performed by Fu and Cohen~\cite{Fu} on BaTiO$_3$ suggest
that the R-T path should be the most energetically favorable one.
This has turned out not to be the case as instead the system crosses
over from the R-T to the O-T path indicated in Fig.~1, resulting in
the appearance of the M$_{\rm{C}}$ phase.  
In addition, the first principles calculations of Bellaiche, Garcia, and Vanderbild predict the R-M$_{\rm A}$-M$_{\rm C}$-T transformation to be reversible upon removal of the electric field~\cite{Bellaiche}. This is not observed experimentally in PZN-8\%PT.


The electric field dependence of the lattice strain of a poled
PZN-8\%PT crystal measured along the field direction has also been
studied.  The sharp jump in the $c$-axis lattice spacing previously
observed with x-rays was not reproduced by our neutron experiment.
Instead, a marked asymmetry of the (002) Bragg peak line shapes
measured along the $c$-axis is observed, and is ascribed to a
non-uniform strain distribution within the crystal that in turn leads
to distributed $c$-axis jumps throughout the crystal volume to the
high-field tetragonal (T) phase.  These results are consistent with
previous experimental observations using low and high-energy x-rays,
and demonstrate the important finding that the responses of the skin
(surface) and bulk (inner volume) regions of the crystal are different
in these piezoelectric materials.

\section*{Acknowledgments}

We would like to thank 
D.\ E.\ Cox, 
L.\ E.\ Cross, 
K.\ Fujishiro, 
A.\ Fujiwara, 
M.\ Matsuda, 
M.\ Nishi, 
Y.\ Uesu 
D.\ Vanderbilt and 
Z.\ Zhong, 
for stimulating discussions. 
This work was supported in part by a Grant-In-Aid for Scientific Research 
from Japanese MONBU-KAGAKUSHO, and by the U.\ S.\ Department of Energy under 
contract No. DE-AC02-98CH10886.  We acknowledge the support of the NIST 
Center for Neutron Research, the U.\ S.\ Department of Commerce, for providing 
some of the neutron facilities used in this work.  This study was supported mainly by the U.\ S.\ -Japan Cooperative Research Program on Neutron Scattering between USDOE and MONBU-KAGAKUSHO and partly by the RIKEN NOP Project.

\end{document}